\documentclass[epj]{webofc}
\usepackage[varg]{txfonts}  

\woctitle{CHARM 2018}
\begin{document}

\title{X, Y, Z Search at Belle II}
%
%

\author{\firstname{Elisabetta} \lastname{Prencipe}\inst{1}$^,$\thanks{\email{e.prencipe@fz-juelich.de}}, on behalf of the Belle II Collaboration}

\institute{Forschungszentrum J\"ulich (DE)}

\abstract{Search for exotics has increased importance since the observation of the X(3872), 13 years ago, announced by the Belle Collaboration. The observation of pentaquark states by LHCb, and the Z-charged states observed at Belle  and BES III have raised even more the attention to the field. Presently several states are observed that do not fit potential models, and looking for them in different production mechanisms and search for their decay modes it is important, as well as to do precise measurement of their mass, width, lineshape. We shortly report in this note about the plan in searching for exotics at Belle II at KEK (Tsukuba, Japan), that just ended the Phase-II running period, and show the first re-discovery results using 5 pb$^{-1}$ integrated luminosity. 
}

\maketitle

\section{Introduction}
\label{intro}
The Gell-Mann Zweig idea, known as the Costituent Quark Model~\cite{eli1, eli2} (CQM), classifies all known hadrons, and is still valid after more than half century, although in the meantime the quark $top$ and $bottom$ have been added to the pattern rising the number of quark families to 3. QCD- (Quantum Chromodynamics) motivated models based on this idea predict the existence of more complex structures than simple mesons (2 quark-bound states) or baryons (3 quark-bound states). Since the observation of the $X(3872) \rightarrow J/\psi \pi^+ \pi^-$~\cite{eli2bis}, and subsequent confirmations~\cite{eli3, eli4, eli5, eli6, eli7, eli8, eli8bis} it became evident that the Charmonium-like states with more than 3 quarks exist, and a flood of new theories and possible interpretations have been brought up.

Until 2003 theoretical models and experiments have got an overall agreement of 2-3 MeV/$c^2$ precision in the mass measurements of Charmonium states. After the observation of the X(3872) in 2003, from experimental point of view, lots of effort was put in searching for new forms of bound states. Some of them well fits the theory, while others are found at unexpected mass values, and still several possible interpretations are opened, mostly due to the poor available statistics in the past experiments to investigate them,  $i.e.$ performing  a full amplitude analysis, or detector limitations did not allow to measure the very narrow width, crucial to discriminate among different theoretical interpretations~\cite{elinora}. Undoubtedly, the  Belle experiment~\cite{elibelle} gave an important and substantial contribution to the field.

Meson, baryons, dibaryons, glueballs, hybrids, tetraquarks, pentaquarks are only some of the hypoteses  populating the Charmonium-like literature. An upgrade of the Belle experiment is then needed to go beyond the spectroscopy achievements of the Belle experiment, and try to give answers to still open questions.

\section{The Belle II experiment}
The Belle II experiment is located in Tsukuba, Japan, at the KEK facility where the old Belle experiment was located~\cite{eli17}. It is a huge international collaboration of roughly 800 physicists, 25 countries. The experiment structure is similar to that of the previous Belle experiment, with few substantial differences listed here below:
\begin{itemize}
\item a Pixel detector (PXD) has been planned, with vertex resolution in z-direction a factor 2 better than at Belle: from 50 $\mu$m (Belle) to 25 $\mu$m (Belle II);
\item the Time-of-Propagation (TOP) detector has been installed and properly working for particle identification purposes. The time resolution is 50 ps by design, and the detector surface is polished at nanometer precision;
\item KLM detector for $K_L$ and muon detection: 2 inner layers of barrel and all layers in the endcap are replaced by scintillators;
\item the electromagnetic calorimeter (ECL) readout electronics has been exchanged: now fast ADCs are used;
\item a gain-factor 40 better than Belle luminosity is planned, due to the new value of the beam current (which gives a factor 2) and the nano-beam principle, which gives an improvement of a  factor 20. In this way we expect 50 times more data than what was collected at Belle over 11 years, so in 2026 the recorded integrated luminosity is expected to be 50 ab$^{-1}$.
 
  \end{itemize}

\section{Y-vector family search at Belle II}
\label{sec-1}
The search for resonant states via Inistial State Radiation (ISR) production started in BaBar, in 2005, with the first observation of the Y(4260)~\cite{Ybabar}. Several other states were then observed by analyzing the invariant mass distribution of $J\psi \pi^+ \pi^-$ and $\psi(2S) \pi^+ \pi^-$ in processes like $e^+e^- \rightarrow \gamma_{ISR}J/\psi$($\psi(2S)$)$ \pi^+ \pi^-$. The quantum numbers of these states are obviously known, $J^{PC}$ =1$^{--}$, but still not clear their interpretation. The Y(4260) was then confirmed by Belle~\cite{Ybelle} in the invariant mass system of $J\psi \pi^+ \pi^-$, where the Belle collaboration claims also the evidence of a new state, the Y(4008); then the Belle collaboartion announced also the Y(4360) and Y(4660) in the invariant mass system of   $\psi(2S) \pi^+ \pi^-$ via ISR processes~\cite{Ybelle2}. This search is peculiar in $e^+e^-$ detectors. In particular, Belle II will be in a unique position to perform such analyses with 50 ab$^{-1}$, which was suffering for statistics limitation at the previous BaBar and Belle. It is interesting to notice that all the observed Y-vector states range in roughly 1 GeV/$c^2$ mass window, but they do not mix, which is surprising. Further investigation is then needed with the huge Belle II data set.
The new nomenclature in the PDG~\cite{pdg} indicates these Y-vector states as $\psi$-states.

\section{Z-charged search at Belle II}
The first Z-charged charmonium-like state was observed at Belle~\cite{eli10, eli10bis}, then confirmed later by LHCb~\cite{eli11}. Z states are clearly exotics, as they are charged, while Charmonium states are supposed to be neutral. In the past years several Z states were observed at the BES III experiment, in $\pi^+ J/\psi$ ($e.g.$ the Z(3900)~\cite{eli12}), or $\pi^+ \psi$' ($e.g.$ the Z(4050)~\cite{eli12bis}). It is interesting to notice that all these $exotic$ observations have come over almost the full available Belle statistics, $\approx$ 1 ab$^{-1}$, and in some case Belle obtained only the evidence of such states. This implies that further search is needed, with improved statistics. For most of these states quantum numbers are still not properly defined, due to the lack of statistics that made not possible an angular analysis. Our understanding is that the Z-charged states can be classified in 2 main categories: those quite large, apparently not connected to threshold ($e.g.$ the Z(4430) and the Z(4200), seen in B meson decays); and those pretty narrow, directly seen in $e^+e^-$ collisions and/or ISR processes, such as the Z(3900) and the Z(4020). The Belle II experiment with the planned huge statistics can continue the search in this field, and it is in a unique position, as it can search for both types of Z-charged states, and eventually find the connection and understand the pattern.

 Definitively more investigation with higher statistics  is needed in  this field.

\section{Charmonium in ISR: perspectives at Belle II}
The plan with the first years of data taking at Belle II is to:
\begin{itemize}
\item study the linehape of the Y(4260);
\item try to find $strange$ partners of the Z(3900), $e.g.$ studying the invariatn mass system of $K^+ J/\psi$;
\item evaluate cross sections of exclusive $c \bar c$+hadrons
\item study radiative decay of the X(3872)
  \item study exotics in the $D^* D^{(*)}$ invariant mass system, now possible because of the expected better slow pion detection compared to Belle II. 
\end{itemize}

\section{Bottomonium perspectives at Belle II}
Bottomonium spectrum is significantly different from Charmonium spectrum. This field represents a challenge that Belle II can take, exploring the search of $Z_b$ states, which up to now were observed only in $\Upsilon(5S)$ decays. SuperKEKB can reach the energy in the center of mass of $\approx$11 GeV, meaning that the $\Upsilon(6S)$ decays will be accessible. It represents a unique possibility for the Belle II experiment. The plan is to look then for radiative transitions between bottomonium states, not possible at the previous Belle experiment due to the lack of statistics. The current results on $Z_b$ search are well summarized in Refs.~\cite{Zbbelle, Zbbelle2}, where the branching fractions of $Z_b(10610)$ and $Z_b(10650)$ are reported. Belle II can take a big advantage in this search for the future: if $Z_b$ are loosely bound states, several new molecular states should show up in the spectrum. We plan to:
\begin{itemize}
\item look for conventional bottomonium states in $\Upsilon(5S)$ and $\Upsilon(6S)$ decays, searching for predicted and still non-observed resonant states, via single transitions and double cascade; and filling the remaining spectrum to measure the effect of the coupled channel contribution.
\item search for new exotics in $\Upsilon(5S)$ and $\Upsilon(6)$ decays. With the first 100 fb$^{-1}$ we will take an exploratory physics run at the $\Upsilon(6S)$ energy in the center of mass; when Belle II will reach 1 ab$^{-1}$ integrated luminosity at the in $\Upsilon(5S)$ energy in the center of mass the search for exotics could deliver the first results.
  \item scan the $\Upsilon(5S)$ and $\Upsilon(6)$ mass. It is known that the $\Upsilon(5S)$ and the $\Upsilon(6)$ behave differently when analyzing the $\pi^+ \pi^- \Upsilon$ and $\pi^+ \pi^- h$, which gives hint of a non-$bb$ nature of the $\Upsilon(5S)$. Investigating extra resonances, $e.g.$ at the mass value of 10750 MeV/$c^2$, it is a mandatory request. 
  
\end{itemize}

With the high statistics planned at Belle II we have also an approved plan related to the $\Upsilon(3S)$ analyses:
\begin{itemize}
\item exotics in transitions. The analysis of $\Upsilon(3S) \rightarrow \pi^+ \pi^- \Upsilon(1S, 2S)$ is still limited by statistics. With the expected data to collect at Belle II, full amplitude analysis will be possible, and search for missing dipion $\pi^+ \pi^-$- and/or $h$-transitions to further constraint theoretical models as well. Radiative transitions represent a unique possibility at Belle II.
\item search for Charmonia in production. With the first expected data set of 300 fb$^{-1}$ integrated luminosity, Belle II will obtain x5 higher sensitivity in inclusive $\Upsilon(3S)$ production, and x15 in double charmonium production. The inclusive rate of the X(3872) could be accessible, and testing the nature of the X(3872) in $\bar D D^*$ as well.
\item analyzing rare $\chi_c$ decays will be possible;
  \item deuteron production mechanism will be under investigation.
  \end{itemize}

\begin{figure}[ht] 
\centering
\includegraphics[width=6cm]{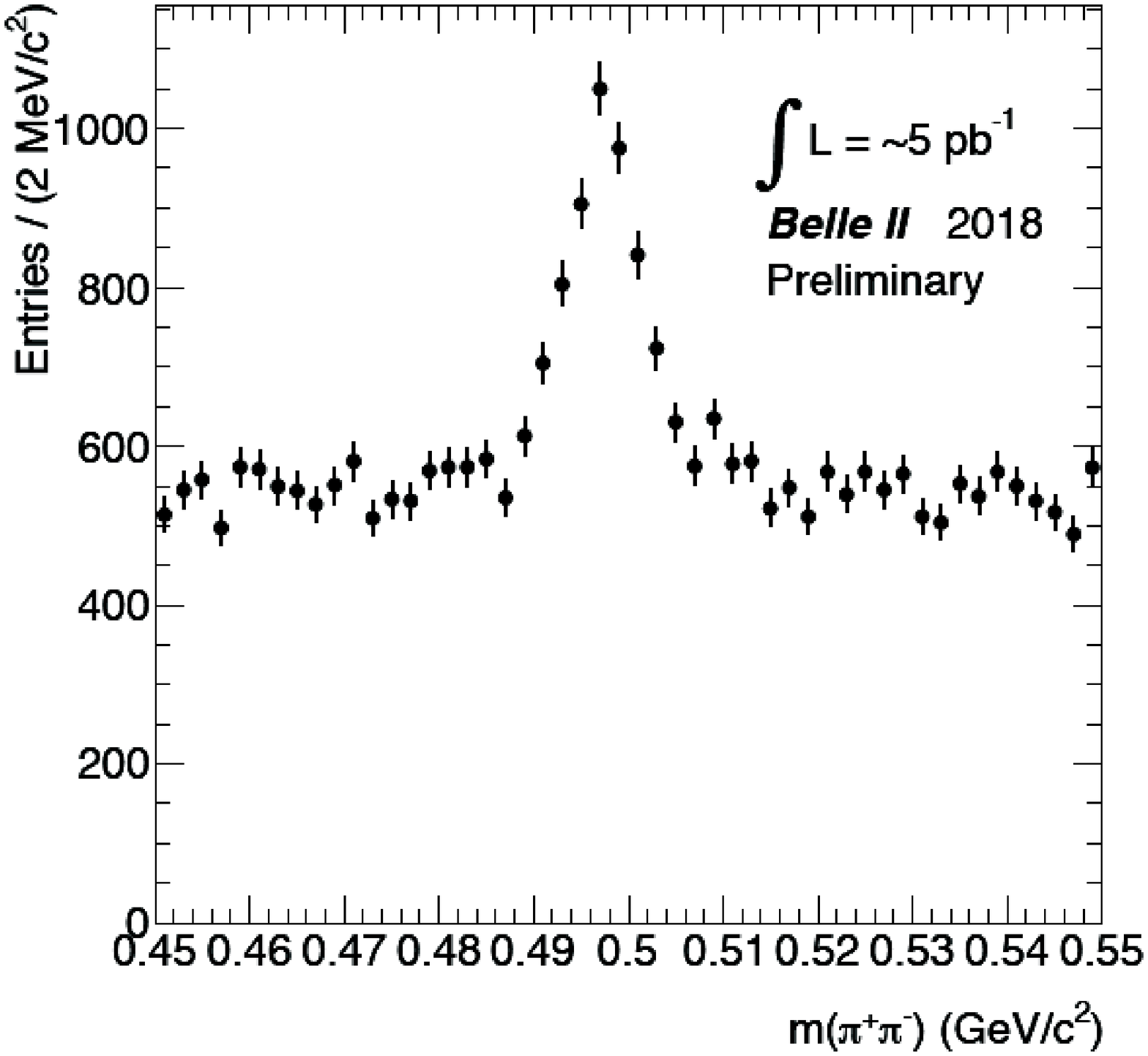}
\includegraphics[width=6cm]{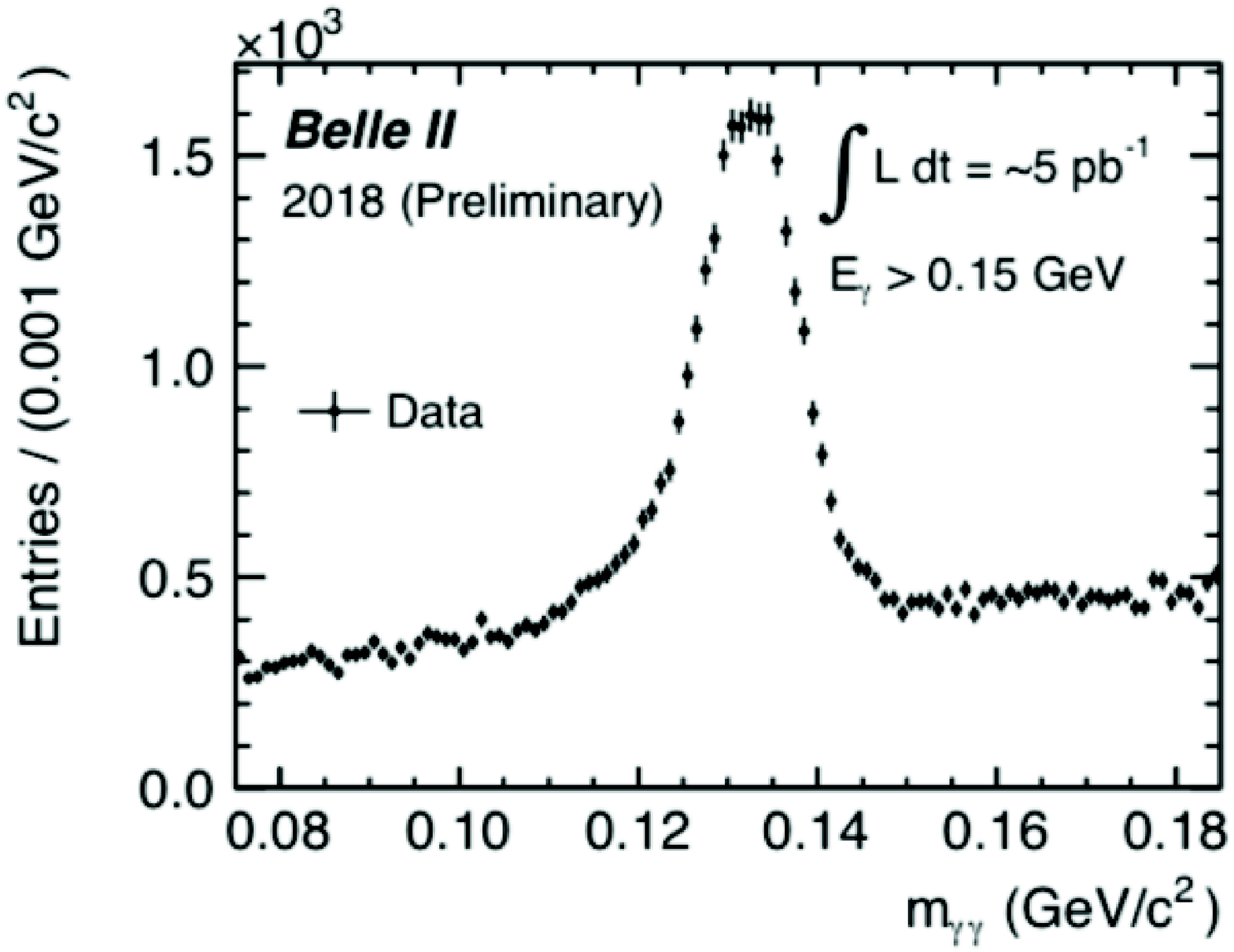}
\caption{ \label{fig1}(left) Preliminary plot of the invariant mass distribution of the $\pi^+ \pi^-$ system on 5 pb$^{-1}$ integrated luminosity with Belle II data set. A clear peak is shown at the $K^0_s$ nominal mass; (right)  invariant mass distribution of the $\gamma \gamma$ system on 5 pb$^{-1}$ integrated luminosity with Belle II data set. A clear peak is shown at the $\pi^0$ nominal mass, after applying a cut on the photon energy larger than 150 MeV.}
\end{figure}
\begin{figure}[ht] 
\centering
\includegraphics[width=6cm]{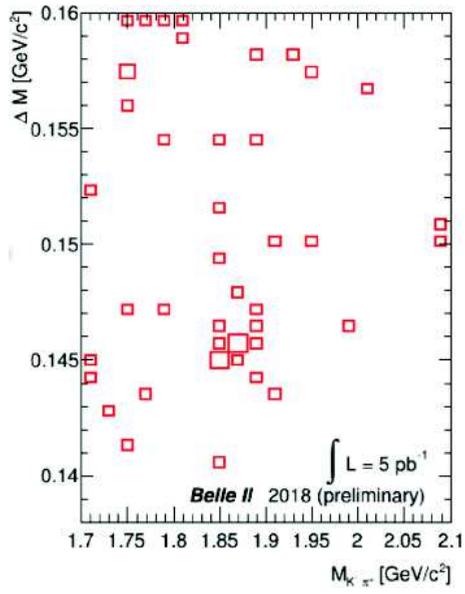}
\caption{ \label{fig2} Preliminary plot of the invariant mass distribution of the $K^- \pi^+$ system vs the mass difference $D^* - D^0$ in the decay $D^* \rightarrow D^0 \pi$, $D^0 \rightarrow K^- \pi^+$, on 5 pb$^{-1}$ integrated luminosity, with Belle II data set.}

\end{figure}

\subsection{Summary}
The Belle II experiment started the Phase II of data taking on 14$^{th}$ of February, 2018. Cosmics were collected to check the detector response and for alignment purpose. Later on, on 26$^{th}$ of April, 2018 the first luminosity run started, and first collisions happened, showing that the experiment is in good shape. In Figs.~1-2 first results on data are shown: the $\gamma \gamma$ invariant mass and the $\pi^+ \pi^-$ invariant mass distributions are shown over 5 pb$^{-1}$ integrated luminosity. Even with so low luminosity run, hints of charm-physics can be seen, in the reconstruction of the $D^* \rightarrow D^0 \pi$ channel. We are looking forward to see the results of analyses at the end of Phase-II, which just happened on 17$^{th}$ of July, 2018. Phase III will start from next year 2019, when the Pixel Detector will be finally included in the full detector set up. Great achievements are expected when Belle II will collect the first 10 ab$^{-1}$, planned by the end of 2020. 

\clearpage


\end{document}